# On the possible origin and evolution of genetic coding

Jacques H. Daniel (1) (2)

((1) Centre National de la Recherche Scientifique, Centre de Génétique Moléculaire, Gif-sur-Yvette, France, (2) InvenTsion, Rehovot, Israel, as present address; Correspondence:jacques.daniel1@gmail.com)

To synthesize peptides alongside the RNAs, some genetic coding involving RNA had to develop. Herein, it is proposed that the first real-coding setup was a *direct* one, made up of continuous poly-tRNA-like molecules, with each tRNA-like moiety carrying, beyond and near its 5' or 3' end, a trinucleotide site for specific amino acid binding: the sequence and continuity of the tRNA moieties of a particular poly-tRNA would ensure the sequence and continuity of the amino acids of the corresponding peptide or small protein. In parallel with these particular entities, and enhancing their peptide-forming function, a proto-ribosome and primitive amino acid-activation system would develop. At some stage, one critical innovation would be the appearance of RNA fragments that could tighten several adjacent tRNA moieties together on a particular poly-tRNA molecule, by pairing with the second trinucleotide sequence (identical to the first one carrying the specific amino acid-binding site) situated at, or close to, the middle of each tRNA moiety (i.e., the present "anticodon"site). These RNA fragments, acting as authentic co-ribozymes in the peptide-synthesizing apparatus, would constitute the ancestors of the present mRNAs. Later, on these mRNA-like guiding fragments, free tRNA forms would be additionally used, first keeping their amino acid-binding sites, then losing them in favor of a specific amino acid attachment at a –CCA arm at their 3' end. Finally, these latter mechanisms would progressively prevail, leading to the modern and universal *indirect* genetic coding system. Experimental and theoretical arguments are presented and discussed in favor of such a scenario for the origin and evolution of genetic coding.

## Introduction

If general coding systems prevail as a result of human's abstract and imaginative thinking, the origin of genetic coding–at the basis of the functioning of every living organism–still remains an unresolved mystery for the scientist.  Obviously, to approach this difficult issue, one should retain the generally accepted principles of (bio)chemistry while attempting to provide a reasonable account of the evolutionary steps required to arrive from a primitive coding system to the elaborate and universal one existing today.

Several theories on the origin of the genetic code have been proposed, which, for simplicity, can be divided into two groups. Some of the theories In the first group account for codon sequence similarity in relation to amino acid family (in particular in relation to amino acid "polar requirement"; Woese et al. 1966) by evoking its effect in minimizing the deleterious outcome of mutations or translation errors (Haig and Hurst 1991; Freeland and Hurst 1998; Freeland et al. 2000); some others suggest that the specific codons co-evolved with amino acid metabolism during primitive evolution (Wong 1975; Di Giulio 1999); yet others even propose that codon assignations were specified in a partially, or even totally, arbitrary manner (Crick 1968; Ohnishi et al. 2002). We could define all of the theories in this group as formalistic speculations because they give no clue whatsoever as to the underlying mechanisms that fashioned the final coding system. The situation differs with the second group, the so-called stereochemical theories: these attempt to envision how the physical interactions between nucleic acids and specific amino acids might have shaped the coding apparatus, at least in its initial stages (Shimuzu 1995; Yarus et al. 2009). Indeed, Michael Yarus, one of the main proponents of stereochemical theory, has come up with some rather precise–although not straightforward–machinery for the primitive coding of small peptides (Yarus et al. 2009). Despite the fact that this second group of theories does address the issue of the origin of genetic coding mechanisms−and not simply the properties and possible evolution of the genetic code from a rather abstract perspective−it seems that new theoretical approaches are required to tentatively arrive at a more realistic and definitive vision of the initial formation and further developments of the genetic coding apparatus during the very early stage of life evolution. Herein, one possible model is presented, together with various compatible experimental results and theoretical considerations.

## A new model for the origin and evolution of genetic coding

Since the main issue of general coding is how to produce and maintain a strict co-parallelism between one sequence of objects, or events, and another (i.e., the encoded sequence), the proposed model postulates the existence of a *direct* mechanism for the very initial genetic coding, which would later evolve into the extant general three-pronged *indirect* system, i.e., from codons to amino acids via free tRNAs (Fig. 1).

1. *Continuous poly-tRNA*. It is suggested that at the very start, some continuous poly-tRNA molecules would guide the direct buildup of peptides, or small proteins. For this, each tRNA moiety of the linear poly-tRNA molecule would carry, beyond and near its 5' or, alternatively, 3' end, a main binding site for some specific amino acid, with no interruption of the continuity of the poly-tRNA entity. Thus, the sequence and continuity of the tRNA moieties of a particular poly-tRNA would ensure the sequence and continuity of amino acids of the corresponding peptide or small protein.

2. *Proto-ribosome and proto-activation of amino acids.* At about the same time, a primordial all-RNA ribosome would appear and develop around the tRNA moieties of the poly-tRNA, and soon, peptides that would improve the functioning of the ribosome and the activation of amino acids would be synthesized from this nascent machinery.

3. *Co-ribozyme*. One critical innovation would be the appearance of RNA fragments that could squeeze together several adjacent tRNA moieties on a particular poly-tRNA molecule, by pairing with the second trinucleotide sequence (identical to the first carrying the specific amino acid-binding site) situated at, or close to, the middle of each tRNA moiety (i.e., the present "anticodon"site). These RNA fragments would further result in significant enhancement of the peptide-synthesizing capacity of the poly-tRNAs and thus, would be actively "selected" (or enriched) for. These RNA fragments, acting as authentic co-ribozymes in the peptide-synthesizing apparatus, would constitute the ancestors of the present mRNAs.

4. *Involvement of free tRNAs*. In this context of rather efficient ribosomal entities and co-ribozyme RNA fragments, a real breakthrough would occur, paving the way for modern genetic coding: the peptide-synthesizing machinery's ability to engage free tRNA molecules, together with the tRNA moieties of poly-tRNAs, with the resulting continuity of the encoded "message" being ensured *in fine* by the mRNA-like fragments. First, the free tRNAs would also include their specific amino acid-binding site at, or beyond, their 5' or 3' end. Later, a new and general mechanism (i.e., the –CCA arm at the 3'end) would be created that covalently links the specific amino acid to each tRNA, stripping its amino acid-binding site sequence, which would have become obsolete.

5. *Triumph of modern translation*. In the "competition" between the tRNA moiety- and free tRNA-based systems of peptide synthesis, the latter would progressively prevail; eventually, with the replacement of RNA by DNA as the genetic memory depository, it would become the exclusive translation mechanism for every living organism on this planet.

## Rationale

Nucleic acids such as RNAs brought an essential characteristic to the chemistry of the primordial soup by allowing the formation and replication of long and diversified chains–rare objects of order and continuity in an otherwise erratic surrounding medium. These quasi one-dimensional entities, which could multiply rather rapidly, would acquire another critical property by popping out chain loops that would anchor firmly, at their two ends, to paired double chains (stems) to produce the so-called hairpins–a kind of conquest of the second dimension (Briones et al. 2009). But the more remarkable achievement would ultimately be the formation of the tRNA-like structure, composed of an aggregation of two such hairpins to form, by self-folding, a real tridimensional molecular object with some conformational rigidity–an L-shaped structure having a rather solid elbow and two arms maintained at distance (Di Giulio 1995; Widmann et al. 2005; Fujishima et al. 2008). The binding of RNA with surrounding small molecules–in particular amino acids–might have already occurred at the hairpin stage since the loop structure would display relatively specific binding sites that were open to the medium components (Shimizu 1995; Rodin et al. 2011). However, to organize these amino acids into a linear chain, or peptide, a more rigid structure for amino acid binding would have been required, offered only by the next structural level brought about by the tRNA pattern. In fact, there would be a need not only for specific binding to amino acids but also for producing some favorable contacts between two amino acid-charged tRNA components, which would enable the creation of a peptide link between their two consecutive amino acids (the tRNA elbow might have been an essential structure for this, exemplified by its binding to the present ribosome at three different ribosomal sites; Zhang and Ferré-D'Amaré 2016). Although such a mechanism may have been relatively efficient in executing the peptide-formation activity, two serious issues would remain to be solved: (i) how to produce longer molecules than bi-or tri-peptides, and (ii) how to make them reproducibly. Here, the RNA's basic features of linear order and continuity would come into play: by providing a template for tRNA anchoring, long sequential stretches of active tRNAs could then be made and replicated, thus allowing for the formation of much longer peptides and constituting a retaining memory of "successful" trials.

In this scenario so far, we deal only with *direct* genetic coding that resembles typical chemical reactions, albeit with the peculiarity that the process of peptide polymerization has somewhat hitched a ride on the RNA polymer organization. Added to this was the increasing sophistication of the ribosome as a scaffold for the peptide reaction (first made of RNAs and later assisted by peptides; Nissen et al. 2000), and of the amino acid-activation system, possibly starting with some multi-competent RNA molecule (Suga et al. 2011) and later replaced by peptides (at the origin of the aminoacyl-tRNA synthetases). However, from the point of view of genetic coding, a breakthrough would begin to surface with the appearance of a "discreet" co-ribozyme activity involving a continuous RNA pairing with the "anticodon" sequences of sequential tRNA moieties of the peptide-coding poly-tRNA. First, this co-ribozyme RNA would help maintain two successive tRNA moieties in the best relative positions to achieve a peptide link effectively; later, this RNA would take a leading role in the coding process, being "read" by charged tRNAs that had become independent entities, fully recyclable and operating at multiple locations. The memory of the system would progressively leave the poly-tRNA molecules in favor of these RNAs, which would become messenger RNAs (mRNAs): an *indirect* genetic coding would thus generally prevail and be used universally and exclusively in all future life forms.

**Possible relics of ancient active poly-tRNAs in today's genetic material**

The tRNA backbone is certainly among the most ancient molecules, with potentially few changes since its first appearance billions of years ago (Byrne et al. 2010). It seems, then, reasonable to question whether some remnant of tRNAs' organization prior to the last universal common ancestor might exist in today's genetic material.

Indeed, it was found in *Bacillus subtilis* that most of the genes encoding the various tRNAs used for translation are not dispersed, but rather clustered within two regions of the genome, each region giving rise to a multigenic transcription product (poly-tRNA mRNA) (Wawrousek et al. 1984). When the two theoretical peptides were derived from these two continuous poly-tRNAs, an interesting result emerged: one peptide presented homology with a whole group of important and possibly ancient proteins, among them an aminoacyl-tRNA synthetase; the second peptide was homologous to several other families of proteins, also of potentially ancient origin. Moreover, these intriguing observations appeared to be corroborated by the existence of rather good homology between the mRNAs encoding these proteins and the putative mRNAs that would encode these putative peptides (Ohnishi 1993; Ohnishi et al. 2005).The authors concluded that the individual tRNAs comprising the poly-tRNAs were directing the synthesis of the two peptides, basically using a ribosomal apparatus and mechanisms similar to those existing today for translation, in particular with two "mRNAs" that would have been created (from two tRNAs) by progressive adaptation and selection of their "codons" against the ordered "anticodons" of the corresponding poly-tRNAs (Ohnishi et al. 2002). However, we find the above scenario complex and rather unlikely. Instead, we believe that these two poly-tRNAs discovered in *B. subtilis*−which have striking equivalences in other types of bacteria such as *Escherichia coli* (Ohnishi et al. 2000)−might represent relics of an older mechanism of translation which directly attached a specific amino acid to a specific tRNA moiety of the full poly-tRNA , leading to the creation of a peptide bond with the neighboring amino acids, the continuity and amino acid order of the resulting peptide being provided by the continuity and tRNA order of the originator poly-tRNA.

Another group of likely most ancient molecules, appearing at the very beginning of the RNA/peptide world, is the ribosomal RNAs (rRNAs). Remarkably, evidence has been presented showing that the entire set of tRNAs for the 20 amino acids were once encoded in both the 16S and 23S rRNAs of the bacterium *E. coli* K12 (Root-Bernstein and Root-Bernstein 2015). Moreover, these tRNA remnants are not found scattered within these rRNA sequences but rather displayed in four different poly-tRNA forms. This intriguing discovery may attest to an ancient mechanism of translation similar to that postulated above−and indeed, for the longest remnant poly-tRNA sequence, for instance, one can detect by BLAST search a derived peptide sequence homologous to a sequence of bacterial aminotransferases or a non-ribosomal peptide synthetase (not shown)−nevertheless, conclusions are harder to draw since we are dealing with putative and remnant–rather than actual−tRNA sequences.

## The tRNA clusters in the *B. subtilis* genome

The new model of genetic coding presented above postulates the existence of main amino acid-specific binding sites beyond, and close to, either the 5’ or 3’ end of the backbone sequence of each tRNA moiety of the putative primitive poly-tRNA molecules.

Obviously, this assumption dismisses, at the dawn of genetic coding, the need for the 5’-CCA-3’ arm found at the 3’ end of all of today’s functional tRNAs and serving as the activated amino acid-acceptor site (this is in contrast to Rodin et al. 2011; Bernhardt and Tate 2010; and many others). However, this should not be viewed as a drastic and rather unlikely hypothesis since, as can be observed in several tRNA genes in the extant genomes (and also among the tRNA genes of the *B. subtilis* clusters), the CCA arm is not encoded in the gene itself but instead, is added later to the formed tRNAs by an enzymatic mechanism (Hou 2010). As a matter of fact, we believe that the primitive “tRNA genes” were all lacking the CCA-encoding sequence at their 3’ end, and it is only later in evolution that most of these genes acquired the CCA sequence by reverse transcription from CCA-added tRNA templates. This view certainly seems more reasonable than the alternative claim that assumes that the precise CCA sequence was deleted in some of the tRNA genes, at some time and for some unknown reason.

Since, as mentioned above, the two tRNA clusters might be of ancient origin, it seemed interesting to compare the various tRNA nucleotide sequences in an attempt to generate some type of “genealogical” tree between these different tRNA species. This seemed to be a critical endeavor, since it might reveal interesting clues on the order and time of appearance of these various tRNAs during evolution–and thus, on the order and timing of the first use of the corresponding amino acids by the translation machinery–as well as possibly other clues on the nature of the translation apparatus itself, or on the state and evolution of the genetic code, for example.

Indeed, the pairwise sequence comparisons of the tRNAs (performed on nearly 40 % of all of the 300 possible combinations and presented in Fig. 2; see Fig. 12 for general codon table and “codons” involved in this particular search) showed that they could unambiguously be arranged into a tree form (Figs. 3 and 4), where each vertical branch length is proportional to the “dissimilarity index” found between two successive tRNAs (as defined in the legend of Fig. 3). Remarkably, all tRNAs appeared to “descend” from only two “ancestors”, the Val-tRNA and the Tyr-tRNA. Tyr-tRNA has a few “descendants”, giving rise only to the three Leu-tRNAs. All other tRNAs “come” from Val-tRNA after one or several “generations” (up to three). It is important not to take the term descendance, or related, as used here, too strictly: it actually denotes a close relationship by a putative mechanism that will be discussed below. Suffice it to say at this point that it is extremely likely that the resulting tree represents some temporal events, related to tRNAs and their cognate amino acids, occurring at the start and during the evolutionary progression of genetic coding.

In this context, it is striking to find nearly all of the tRNAs linked to what are considered the first amino acids occurring in the primeval soup (Higgs and Pudritz 2009), placed at the top of the diagram of Fig. 3: valine, glycine, phenylalanine, proline, alanine, lysine, serine, threonine, isoleucine, aspartic acid, leucine, with the exception of tyrosine (Fig. 3, Box1); and nearly all of the amino acids appearing later, at the bottom of the diagram: histidine, cysteine, tryptophan, methionine, arginine, asparagine, glutamine, with the exception of glutamic acid (Fig. 3, Box2) . These results give even more strength to the interpretation that this “genealogical” tree represents some temporally sequential events that happened in the early evolution of genetic coding. Moreover, they strongly suggest that the first coding systems, developed at the dawn of life, were of an opportunistic nature–using the surrounding amino acids–and only later, when some form of metabolism came out, extending their range of amino acid utilization.

The amino acid-specific aminoacyl-tRNA synthetases−essential intermediates in today’s protein synthesis−are divided into two classes (I and II) with different characteristics, in particular in relation to the tertiary structure of their tRNA-binding site as well as the tRNA acceptor-stem side and groove to which they bind (Artymiuk et al. 1994; Sugiura et al. 2000; de Pouplana and Schimmel 2001). Considering the class of the amino acid involved, an interesting regularity is displayed in the diagram (Fig. 3). After the appearance of the two “ancestor” tRNAs, for valine (presently of class I) and for tyrosine (of class II in bacteria), the great majority of tRNAs in Box1 involve amino acids belonging to class II (9 class II amino acids, with the exception of isoleucine and leucine). However, for Box 2 (taking into account the sole appearance of new amino acids) all amino acids, except histidine, belong to class I. These are not trivial results, as they seem to be telling us something about the evolution of the genetic coding system (see below for possible significance). Moreover, these results, showing that all direct “descendants” of valine belong to class II without exception, raise the possibility that valine was once a class II amino acid which, during early evolution–but before the same switch happened to tyrosine in archaeal bacteria and eukaryotes–became class I. If indeed all of the first amino acids were of class II at the beginning of peptide synthesis, this would mean that the appearance of the primitive class II aminoacyl-tRNA synthetases preceded that of their class I equivalents, in particular, the critical antiparallel β-sheet fold of the class II proteins may have occurred in the course of evolution before the corresponding Rossmann fold found in class I proteins.

What might explain this observed tree, which puts very different amino acids in a close relationship through the sequences of their cognate tRNAs? To answer this difficult issue, a putative primitive sequence was constructed that might somehow be at the origin of the “ancestor” sequences. For this, the nature of the possibly correct nucleotide base was inferred at each position of the tRNA backbone sequence by comparing Tyr-RNA, Val-tRNA and all of its six tRNAs of the first “generation”, and picking the most commonly encountered base. The resulting sequence of this putative molecule, presented in Fig. 5, although coming from several different sequences, was found to retain all of the known characteristics of a *bona fide* tRNA, in particular its proper stems and loops (Fig. 6a).

Notably, this “primitive sequence” can be divided into approximately two halves (made of 33 and 34 nucleotides, respectively) that both display a putative hairpin secondary structure (Fig. 6b; Tanaka and Kikuchi 2001). Also, when comparing the partial sequences of these two halves, there is a rather high, and evenly scattered, level of nucleotide identities (53.6 % on a 28 nucleotide sequence), meaning that these two partial sequences may have derived from a common ancestor (Fig. 7).

We envision such a 33/34 (or possibly 35; see below) nucleotide-long hairpin sequence to be at the origin of all tRNAs. The sequence would first acquire a chain of 3 nucleotides having some affinity for a surrounding amino acid, attached to the hairpin at either its 3’ or 5’ end .The head-to-tail ligation of two such sequences could generate a molecule with a structure close to that of tRNA, with three loops and two identical main amino acid-binding site sequences, one at one extremity of the molecule (see above), the other at the middle loop (the “anticodon”). In this context, it is likely that the actual sequence of the proposed 33-hairpin was 2 nucleotides longer, with two additional adenines at its 5’ end (and thus making it 35 bases long, the two extra adenine bases being inferred from the two adenine bases following the “anticodon”). By the same token, inferring from the two uracil preceding the “anticodon”, it can be postulated that the proposed 34-hairpin would originally have an extra uracil at its 3’ end. In broad agreement with this view, the nucleotide base found just following the 3’ terminus of the tRNA genes of the *B. subtilis* clusters is predominantly a thymine (57 %); as to the bases found before their 5’ terminus, if the major one at position -1 is thymine (73%; not the predicted base in this model, for unknown reasons), at position -2 nevertheless, the expected adenine base is predominantly encountered (54 %).

How, then, can we contemplate the history of tRNA relationships at the dawn of genetic coding? As shown in Fig. 8, there would be one primitive hairpin sequence that could replicate and evolve with time, giving rise to various progeny sequences that would capture, at different times, trinucleotide sequences having specific amino acid-binding properties. As discussed above, ligation of two such (probably) identical sequences would generate the primitive tRNA entity particular to each amino acid. Thus, because in this model there is no co-evolution of the “anticodon” and the rest of the tRNA-generating sequence, but only an unpredictable temporal encounter between the two, the observed quasi-absence of a correlation between tRNA sequences and amino acid types or “anticodons” finds a logical explanation (for exceptions, see below).

The tRNA moieties in the two poly-tRNA clusters of *B. subtilis* are not strictly contiguous, being separated by interspersed regions of different lengths (varying from 3 to 48 nucleotides, with a mean size of 14 nucleotides).  If these clusters are relics of a primordial direct coding mechanism involving the main amino acid-binding sites situated in these interspersed regions, remnants of these sites could potentially be found. Since, according to this theory, each main amino acid-binding site is supposed to have been identical to the corresponding “anticodon” sequence borne on each tRNA, one can look for the specific anticodon sequence, or part of it, outside the whole tRNA sequence, in the vicinity of its 5’ and 3’ ends. More precisely, this search was performed on the nucleotide triplet just before -2 from the 5’ end, and just after +1 at the 3’ end, of each tRNA sequence, in accordance with the proposed construction of a typical tRNA as outlined above. Results were pooled and divided for all of the regions outside the 5’ end and all of the regions outside the 3’ end, in relation to all tRNAs. Moreover, since the various natural amino acids are partitioned into two classes, I and II–as defined by the type of aminoacyl-tRNA synthetase used for amino acid activation–the pooled results were compiled for amino acid class. Notably, after the 3’ end of the tRNA sequences, one finds for class II amino acids, 24 possible nucleotides that are identical to, and at the same position as, 1 nucleotide composing each corresponding anticodon versus only 13 found before the 5’ end (15 nucleotides expected by chance). Furthermore, precisely after the 3’ end (i.e. starting at +2) of these class II amino acid tRNAs, the sequence of the complete anticodon  is observed for one threonine tRNA sequence whereas none is found before the 5’ end. As for the class I amino acid tRNAs, only small and non-significant differences were observed at these specific regions before the 5’ end and after the 3’ end (not shown). Although these results obviously do not constitute rock-solid proof, they are nevertheless compatible with our hypothesis of the existence of anticodon sequences, outside of the tRNA backbone sequence proper, that may serve as specific main amino acid-binding sites. More precisely, they tend to suggest that these main sites are situated after the 3’ end for tRNA moieties linked to class II amino acids and thus, possibly, before the 5’ end for those linked to class I amino acids.

## A broad perspective of the start and evolution of genetic coding

It is rather clear today that the appearance of non-branched polymers such as RNAs afforded the possibility of creating all kinds of new phenomena on this wet planet by combining chemical activities, memory, and room for changes and expansion (Orgel 2004; Szostak et al. 2001). This primitive RNA world, however, also likely “found” its limitations at some point and developed an additional class of non-branched polymers constituted by the peptide/protein series (Alva et al. 2015). How this natural invention, directed by the RNA ancestor system, occurred, and what mechanisms ultimately led to the universal code apparatus existing today, are among the most difficult central questions of evolutionary biology, certainly ranking with the issue of the expansion of biological complexity and its underlying mechanisms (Daniel 2019).

We believe it rather likely that at the very start, physical interactions would have occurred between surrounding amino acids and small RNA sequences, constituting specific binding sites −whatever their strength− of amino acids to RNA.This possibility has received ample confirmation from studies involving the SELEX technique devised to select RNA pieces *in vitro* that carry out their own kind of biological–from binding to enzymatic–activity (Tuerk and Gold 1990; Ellington and Szostak 1990). Amazingly, many such RNAs bearing amino acid-binding sites contained sometimes codon, but mostly anticodon sequences, corresponding to the specific amino acid with high statistical significance (Yarus et al. 2009; Yarus 2017; Rodin et al. 2011). It should be noted that many years before the existence of the SELEX approach, one study among others had arrived at the same conclusions by demonstrating that there is a significant correlation between the properties of amino acids (in particular the polarity and bulkiness of their side chains) and their specific anticodons (Jungck 1978). Moreover, theoretical conformational studies had shown that pentanucleotides containing, in the middle anticodon sequences and at the 5’ end, a uracil base producing a U-turn conformation (no pun intended!), might interact with their cognate amino acids, accounting for “almost all the salient and unique features of the contemporary protein synthesizing machinery”–such as triplet coding, wobble behavior, exclusive use of proteinogenic amino acids, chiral uniqueness, existence of a U-turn conformation in the anticodon loop of tRNA crystal structures (Balasubramanian 1985). All these results, pointing to similar conclusions which do not seem to be trivial, should therefore be taken into consideration when conceiving any theory on the mechanistic origin of genetic coding. We do not know whether at the start of life, this kind of interactions between RNA and amino acids had any benefit, by enhancing some enzymatic activity of the RNAs or by participating in some sort of primitive peptide formation, for instance (Kun et al. 2008; James and Mandal 2011; Szostak 2012; Zhang et al. 2016; Tkaczewska et al. 2019). Nevertheless, it seems that these interactions might have been crucial for generating the universal genetic coding system.

According to the model proposed here, the first stages of this complex achievement would involve hairpin RNA sequences of 35 nucleotides length, which would accommodate a potential main binding sequence of three nucleotides at either their 5’ or 3’ end. Two copies of the resultant RNA would further undergo “ligation” to give a pseudo-tRNA structure, with an “anticodon” at the position seen in modern tRNAs, and a second sequence identical to this “anticodon”at its 5’ or 3’ end. Moreover, to arrive at some peptide-synthesizing machinery with direct RNA–peptide co-parallelism, an additional step would have been required: these pseudo-tRNAs, with various “anticodons”, should have been somehow integrated into a larger linear RNA structure–either by some insertion process into an existing RNA sequence, or by their acquisition of additional nucleotides at both ends followed by mutual ligation–so as to produce poly-tRNA strings with various end-result coding activities.

The direct mechanism of genetic coding postulated here could have been a real breakthrough in the primordial stage of encoded peptide synthesis. An obvious improvement would have occurred with the creation of first, a primitive RNA ribosome that would essentially join together two successive tRNA components of the poly-tRNA so as to increase their probability of forming a peptide bond, and then, peptides or small proteins capable of assisting the nascent ribosomal function and mostly, the incipient aminoacyl-tRNA synthetases. Remarkably, the existence of amino acid-binding sites at either the 5′ end or the 3′ end presumed here, provides a simple and straightforward explanation for the old finding that there are two very different classes (I and II) of aminoaciyl-tRNA synthetases.

Yet, the most impressive creation, with truly revolutionary potential, would have come with the appearance of new RNA molecules being first partially, and then with time fully, complementary to the *putative* sequence resulting from putting all of the anticodons of all of the successive tRNA moieties end to end. These RNAs would have potent co-enzymatic effects by ensuring the precise positioning of the successive tRNA components of the poly-tRNA, and thus, of the bound amino acids for peptide bond formation, as mentioned above. The invention of such assisting RNAs would pave the way for the emergence of the present-day *indirect* genetic coding system where, by becoming mRNAs, they would play a pivotal role in the translation process.

The transition from direct to indirect genetic coding would indeed have been made possible by the existence of these “complementary” RNA molecules, because they would allow free-standing amino acid-binding tRNA structures to be used and inserted into the process of translation in place of the tRNA moieties of poly-RNAs. This would first occur episodically in the void left between two different poly-tRNAs attached to a single assisting RNA molecule, thus reuniting two different peptides into a longer and more complex one; later, it would become the predominant, and then exclusive mode of translation, entering the realm of modern genetic coding. At the same time, the ribosome would become even more complex, in particular by developing unique dynamic properties that would result in directional reading of the mRNAs to give the corresponding peptides.

With the development of free tRNAs participating in peptide formation, there would also be a drastic change in the amino acid-charging mechanism of these tRNA structures. Starting with the presumed main amino acid-binding site at their 3′ or 5′ end, with a sequence identical to the “anticodon”, they would progressively evolve into the modern tRNA form by adding a CCA sequence at their 3′ terminus and covalently linking the specific activated amino acids to it using the aminoacyl-tRNA synthetases.

In relation to the evolution of the genetic coding system envisaged here–with various phases and even profoundly different, though somewhat interwoven, competing mechanisms–it is of interest to interpret the striking results obtained with the genealogical analysis of the tRNAs found here, with its likely implications for defining the historical order of amino acid utilization in peptide synthesis during evolution. As already mentioned, there is an obvious asymmetry between the oldest and first used amino acids and the newest and later used ones, with respect to the class of aminoacyl-tRNA synthetases involved: for most of the former amino acids, the interaction is with class II, whereas the reverse situation is observed for the latter amino acids (Fig. 3). This suggest that at the start of genetic coding, there would have been a strong preference–for unknown reasons–for class II amino acids such that, according to the present model, one end of the primitive tRNA structure would be strongly preferred for attaching the specific amino acid-binding site (possibly the 3’ end; see above); however, later on, conceivably due to the adoption of the free-standing form of tRNAs and possibly the invention of the CCA arm, the preference would have shifted to the other end of the tRNA. Thus, the class type dominance of amino acids used historically may testify to critical events occurring in the genetic coding mechanisms during evolution.

Why did the indirect coding system finally win over the direct one? In our view, the direct primitive system for peptide synthesis paved the way for the indirect one by promoting the tRNA cloverleaf structure as an essential tool, by building up the ribosome scaffold and, last but not least, by allowing low-profile RNAs, used initially as co-ribozymes, to become mRNAs–the central master piece of the universal indirect coding system. This evolution has certainly had several major benefits. One of them is broad economy of means: instead of “freezing” one tRNA for each amino acid used in any peptide within a long RNA chain, there is now a storehouse of recyclable tRNA molecules that are sequentially attached, by their anticodon loop, to shorter mRNA molecules. Furthermore, this likely allowed the evolution of a much more efficient, rapid and faithful mechanism for peptide synthesis and, as a critical outcome, afforded the possibility of creating longer peptides and proteins, with more complex functions. Once the indirect coding system became exclusive and all of the possible trinucleotide codons had been assigned (see below), the capacity of this final system to sustain mutational assault would appear rather outstanding: very small, or mostly point mutations–which might have easily ruined the ability to synthesize a particular peptide in the direct system–would now be much less detrimental because they would not generally prevent elongation of the relevant peptide/protein. Moreover, by changing the peptide/protein sequence locally, some of these mutations might even provide an opportunity for radically improving or enriching its function. In conclusion, adopting the indirect genetic coding system would have put nascent life onto a formidably efficient path of growth, complexity and diversity.

It is interesting to consider the formalistic theories of the genetic code in view of the scheme presented here for the possible origin and evolution of the genetic coding apparatus. The genealogical study of the tRNAs revealed that the amino acids thought to be present in the primordial soup were indeed first utilized for peptide formation, apparently before the amino acids resulting from active synthesis by primitive metabolic pathways. Thus, peptide synthesis seems to have preceded amino acid metabolism, which may therefore have been dependent on peptides for its occurrence and full manifestation (alone or with the assistance of RNAs). As a result, one can argue for the plausibility of co-evolution of the genetic code and amino acid metabolism, as proposed long ago by several authors (see Introduction).

The genetic code is thought to be robust (although not among the most robust; Novozhilov et al. 2007), such that it can minimize the effect of single mutations via modification to another codon specifying either the original amino acid, or one in the same chemical family. We attempted to address this issue with respect to the historical events behind the codon-capture phenomenon by looking at tRNA genealogy applied to other tRNAs of *B. subtilis*, i.e., those not encoded in the two large clusters (Fig. 9), as well as to tRNAs of another bacterium, *E. coli* (Fig. 10), all chosen because of the difference between their anticodons and those of the tRNAs of the *B. subtilis* clusters. In Fig. 11, the map of these other tRNAs from *B. subtilis* is added to, and superposed on, the map of the cluster tRNAs (Fig. 3). Remarkably, the codon usage for the involved amino acids (valine, alanine, threonine, leucine and arginine) appears to extend incrementally with time, each step via modification of only one nucleotide in the anticodon sequence. Within the framework proposed here, the simplest explanation might be that a tRNA with such a mutation will keep the same amino acid charged at the other corner of the molecule –i.e., the amino acid-acceptor region, whether in the poly-tRNA matrix, the primitive independent tRNA with an amino acid-specific binding site, or the later tRNA with a CCA arm–thanks to the cognate aminoacyl-tRNA synthetase at each of its stages of evolution. However, in the final genetic code, many amino acids occupy a whole square of the code table without impinging upon other squares (Fig. 12); this seems to suggest that there might have been a 2-nucleotide code *for amino acid binding* (equivalent to positions 2 and 3 in the "anticodon"). Thus, a mutation occurring in the first nucleotide of the amino binding site (equivalent to position 1 in the "anticodon") and appearing at the first stage of the poly-tRNA matrix, or at the appearance of the primitive independent tRNA carrying an amino acid-specific binding site (but no later), along the mechanism described above and in Fig. 8, might better account for the subsequent codon captures by an identical amino acid.

A further codon extension (and confirmation of that delineated above) can be described using the partial complementary tRNA map from *E. coli* (Fig. 11): it concerns serine, proline, threonine, glutamine and glycine. Altogether, in this study using tRNA genealogical maps, 22 codons are left with no anticodon-specific tRNA; however, at the end of such a tRNA's evolution, due to the wobble phenomenon (Crick 1966; Agris et al. 2018), each mRNA could obviously be fully translated, with generally no risk of untimely arrest of the translation process.

Clearly, this description of codon assignment only concerns the ancient/standard genetic code. At apparently much later stages in evolution, and for some species, several minor variations on this theme would occur. Intriguingly, the great majority of these changes seem to involve the class I amino acids and their codons (Ohama et al. 2008).

## Concluding remarks

It is fascinating that at the dawn of life, genetic coding might have evolved from a direct system of correspondence to an indirect and symbolic one, according to the scenario presented here. One can speculate on some basic and general feature that might have allowed such a transition at the origin of all living beings. Somewhat similar to the way in which present cells *duplicate* themselves, either allowing the possibility to produce small modifications in some of their descendants which might then be selected for due to the improvement they provide in a certain environment (competition arena), or alternatively, creating the opportunity to form groups of cells that functionally complement each other (collaboration arena), it is suggested here that a hairpin-like RNA molecule having acquired a small amino acid-specific binding site would have "generated" a *duplicate* form, leading to a tRNA-like structure. The two identical "binding site" sequences of this new structure would have undergone different types of functional evolution: one, being the effective site for binding the cognate amino acid, would later totally disappear following the formation of an alternative and more complex binding system; the second, on the other hand, would perform a unique structural, and then informational role, by specifically pairing with a complementary RNA used first as a co-ribozyme and later as a messenger.

At the start, amino acids present in the primordial soup would have been assembled into small or middle-size peptides that would progressively allow for the metabolism and synthesis of new additional peptide-forming amino acids, as well as the development of a robust ribosomal machinery and coding system, thus triggering a formidable self-amplification process leading to the unique dynamic organization called life.

If one wants to compare this revolutionary transformation in genetic coding that may have occurred in nature to one that took place in human culture, the best instance would probably be the passage from a strictly pictorial representation of reality−as our vision sees it−to a scriptural one−i.e., some writing system transcribing the flow of our speech. Both representations used the same tools–the hands, or some modern equivalent (which could be compared to the basic tRNA structure in genetic coding)–but their potential effects on the cultural evolution of humans were totally different. Only the writing system, in phase with the spoken language, would fully allowed the explosion of ideas, and the exploration of our world's richness, science, technology and art. This would be the equivalent of the profusion of living forms which have been created for billions of years on this planet, with apparently no limit to nature's inspiration.

## Figure legends

**Figure 1**. *Proposed model of the origin and evolution of genetic coding*. Five different coding stages before the advent of modern coding are schematically drawn. Double-lined inverted L-shaped objects, in violet and green, represent primitive tRNAs that are either free, or linked together by spacers to form poly-tRNAs. Red and blue bars represent the nucleotide sequence constituting the amino acid-binding site and "the anticodon site". Oblique unbroken black line represents the RNA co-ribozyme. See text for description and rationale.

**Figure 2**. *Nucleotide sequence comparison of the tRNAs from the two clusters in the* Bacillus subtilis *genome*. Numbers represent Similarity Index (SI; with values from 0 to 100) between two tRNAs (deprived of their eventual CCA 3' end), computed as follows: the number of identical nucleotides given in per-cent of overlap, obtained by running the LALIGN program (at EMBL-EBI; Huang and Miller 1991), was multiplied by the number of overlapping nucleotides, and divided by 73 (i.e. the standard nucleotide length of tRNA). The additional loop sequence occurring in a few tRNAs (for Tyr, Leu and Ser) generally appeared in the comparison as one "pseudo-loop" (or, more correctly, a gap in the facing tRNA sequence or, eventually, two facing "loops"), and the overlapping sequence and division number definitions did not take it (or them) into account, except for intra-Tyr, intra-Leu and intra-Ser comparisons. In these special cases, the sequence overlap of the additional loop was taken into consideration, and the nucleotide length of the additional loop was added to the division factor (i.e., 73) used for computing SI values. The symbol **c** before a value relates to the anticodon complementarity of the two compared tRNAs; the symbol **inv** before a value means that the SI relates to similarity between normal and inverted sequences of the two tRNAs.

**Figure 3**. *Tree arrangement of sequence similarity of the tRNAs from the two clusters in the* Bacillus subtilis *genome*. For a direct link, the length of the vertical line drawn between two tRNAs is proportional to the dissimilarity index, whose value is equal to 100 minus the similarity index (SI). The tRNAs of class I are in black, those of class II in red (corresponding generally to the present tRNA status in bacteria). See Fig. 2 and text.

**Figure 4**. *Average and distribution of tRNA similarity index (SI) values according to type of tRNA link*. SI values are from Fig. 2; direct or indirect links between tRNA are defined by the tree in Fig. 3. For each link category, the vertical bar represents the range, and the circle represents the average, of its SI values. A negative scale for SI is added here to take into account the inverted similarity found in the comparison between some of the indirect tRNA links (see Fig. 2).

**Figure 5**. *Putative primitive sequence that generated tRNAs*. See text for rationale.

**Figure 6**. *Secondary structure of the putative primitive sequence that generated tRNAs (a) and of each of its two halves (b)*. Obtained from the RNAfold WebServer, Institute for Theoretical Chemistry, University of Vienna, Austria.

**Figure 7**. *Sequence comparison between the two halves of the putative primitive sequence that generated tRNAs*. Obtained by LALIGN. See legend to Fig. 2.

**Figure 8**. *Proposed model for generating the various tRNAs*. For each tRNA, and at a certain time during evolution, a hairpin RNA would acquire, at one of its ends, a trinucleotide sequence containing an amino acid-binding site (small horizontal bar in blue, with ”+” sign above or below, it); following duplication (by whatever mechanism) of this augmented hairpin (marked “x2”, with left to right arrow), a primitive tRNA would be produced with a specific amino acid-binding site at one end, and an “anticodon” in the middle, of the molecule (see primitive tRNA at the first stage of genetic coding evolution in Fig. 1). To simplify the drawing, after the second tRNA generation, blue bars and arrows are implicit. At each tRNA position, only one of the two ends of the hairpin accommodated the amino acid-binding site, the same end being used for all the “amino acids“of the same class (exclusion of the other possibility being expressed in the drawing

by parentheses around the other hairpin). For tRNAs with a question mark, another possible mechanism of generation may be at work (see Fig. 11 and text). See text for rationale.

*For Val-tRNA, see text.

**Figure 9**. *Nucleotide sequence comparison of tRNAs for identical amino acids*. Similarity index (SI) was computed, as described in the legend to Fig. 2, by comparing tRNAs for identical amino acids originated from the two clusters, as well as dispersed genes, of *Bacillus subtilis*.

**Figure 10**. *Nucleotide sequence comparison of tRNAs for identical amino acids from* Bacillus subtilis *and* Escherichia coli. Similarity index (SI) was computed as described in the legend to Fig. 2. See text.

**Figure 11**. *Additional tree arrangement of sequence similarity using the tRNAs for identical amino acids from* Bacillus subtilis *and* Escherichia coli. Superposed on the tree arrangement of Fig. 3, are additional links related to the other tRNAs for identical amino acids from *B. subtilis* [at least one not belonging to the two clusters (except for Ser and Leu), and shown as round dotted vertical lines; see Fig. 9], and the tRNAs for identical amino acids from *E. coli* (shown as square dotted oblique lines; see Fig. 10b). See legends to Figs. 2 and 3.

**Figure 12**. *Standard amino acid codon assignment*. Codons in boxes are complementary to the anticodons used by the various tRNAs from the two tRNA clusters of *Bacillus subtilis*. See text.

**Figure 1**

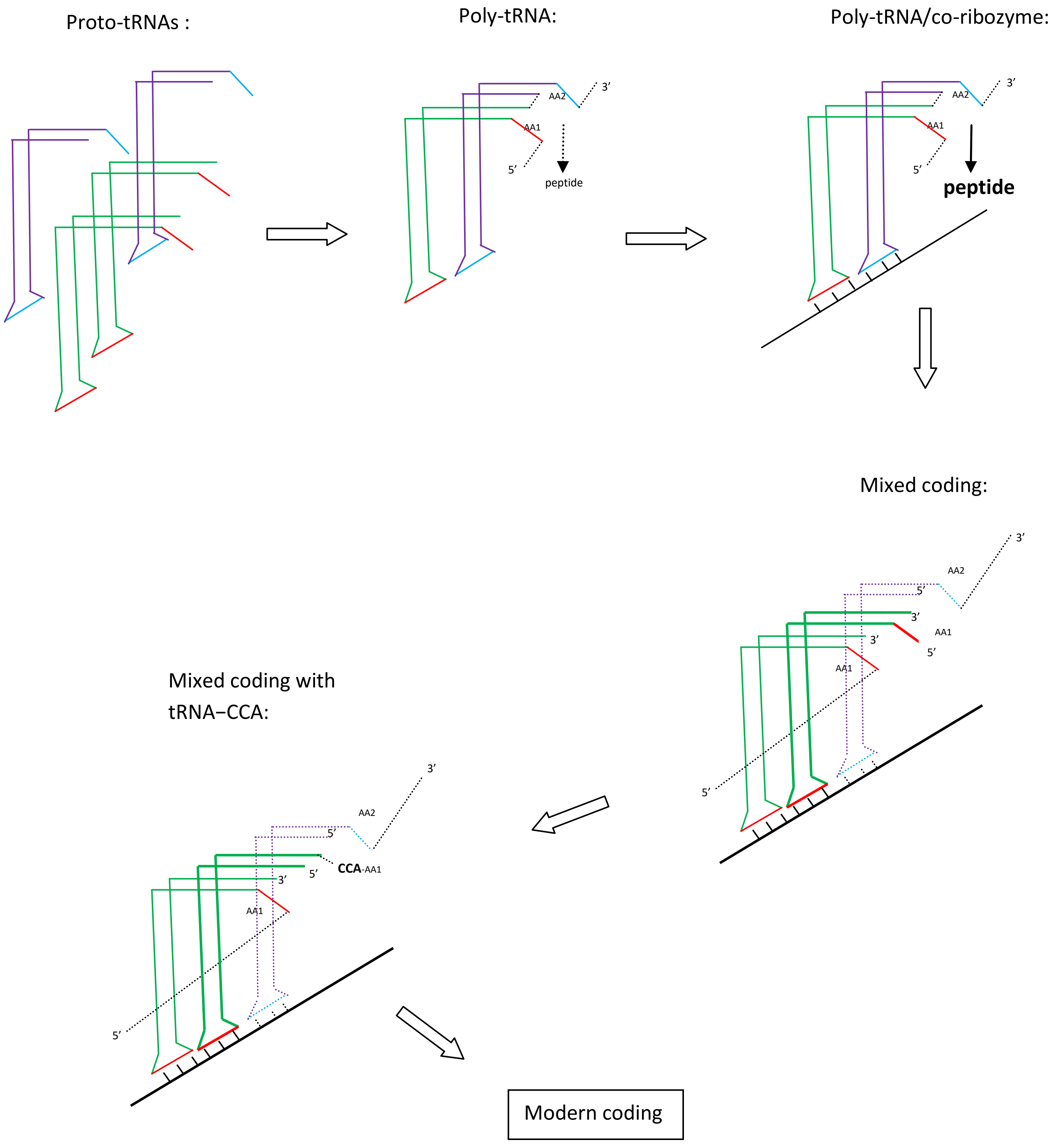

**Figure 2 A**

| | Phe GAA | Leu TAA | Leu CAA | Leu CAG | Ile GAT | Met CAT | Val TAC | | Ser GGA | Ser TGA | Pro TGG | Thr TGT | Ala TGC |
|---|---|---|---|---|---|---|---|---|---|---|---|---|---|
| Phe GAA | | 55.2 | 49,6 | 42.5 | | | 63.2 | | 49.3 | 44.7 | 58.2 | | 50.2 |
| Leu TAA | | | 69.8 | 65.6 | 38.4 | | | | | | 39 | | |
| Leu CAA | | | | 66.7 | inv 36.2 | | 23.5 | | | | | | |
| Leu CAG | | | | | 21.9 | | | | | | | | |
| Ile GAT | | | | | | 69.9 | 56.5 | | | | | | 66.9 |
| Met CAT | | | | | | | 61.7 | | | | | | 62.1 |
| Val TAC | | | | | | | | | | 57.3 | 65.8 | 61.6 | 69.9 |
| | | | | | | | | | | | | | |
| Ser GGA | | | | | | | | | | 62 | | | |
| Ser TGA | | | | | | | | | | | 26 | | 28.7 |
| Pro TGG | | | | | | | | | | | | 63.1 | 61.6 |
| Thr TGT | | | | | | | | | | | | | 57.5 |
| Ala TGC | | | | | | | | | | | | | |
| | | | | | | | | | | | | | |
| Tyr GTA | | | | | | | | | | | | | |
| His GTG | | | | | | | | | | | | | |
| Gln TTG | | | | | | | | | | | | | |
| Asn GTT | | | | | | | | | | | | | |
| Lys TTT | | | | | | | | | | | | | |
| Asp GTC | | | | | | | | | | | | | |
| Glu TTC | | | | | | | | | | | | | |
| | | | | | | | | | | | | | |
| Cys GCA | | | | | | | | | | | | | |
| Trp CCA | | | | | | | | | | | | | |
| Arg ACG | | | | | | | | | | | | | |
| Ser2 GCT | | | | | | | | | | | | | |
| Gly GCC | | | | | | | | | | | | | |
| Gly TCC | | | | | | | | | | | | | |

**Figure 2 B**

| | Tyr GTA | His GTG | Gln TTG | Asn GTT | Lys TTT | Asp GTC | Glu TTC | Cys GCA | Trp CCA | Arg ACG | Ser2 GCT | Gly GCC | Gly TCC | Prim tRNA |
|---|---|---|---|---|---|---|---|---|---|---|---|---|---|---|
| Phe GAA | 55.4 | 47.9 | 53.7 | | 60.6 | | c 52 | 39.1 | | | 36.2 | | | 64.9 |
| Leu TAA | 48.8 | | inv 42.5 | | | | | | | | | | | |
| Leu CAA | 60 | | c inv 42.5 | | | | | | | | | | | |
| Leu CAG | inv 43.2 | | | | | | | | | | | | | |
| Ile GAT | | | 58.4 | | | | | | | | | | | |
| Met CAT | | | 55.8 | | | | | | | | | | | |
| Val TAC | 54.7 | 40.2 | 30.6 | | 70.1 | 56 | 37.5 | 29.3 | | | | 66.2 | | 76.9 |
| Ser GGA | 50.2 | | | | | | | 36.2 | | | 65.6 | 58.9 | c 30.1 | |
| Ser TGA | 53 | inv 44.2 | | | | | | 48.9 | | | 65.6 | 63.5 | inv 38.3 | |
| Pro TGG | 50.1 | 50.5 | 51.7 | | 59.6 | | | 53 | c 60.3 | | | | | 69.5 |
| Thr TGT | | | 57.2 | | | | | | | | | | | |
| Ala TGC | inv 40.5 | 31.5 | 57.7 | | 65.7 | 58.5 | 44.6 | c inv 45.2 | | | | 52.9 | 46.7 | 66.1 |
| Tyr GTA | | 50.7 | | | 44.9 | | | 49.9 | 38.8 | | 41.3 | | | 58.6 |
| His GTG | | | 57.5 | 43.8 | 45.2 | 45.2 | 48.6 | | | 54.8 | | 63.3 | 63.5 | 54.9 |
| Gln TTG | | | | 49.3 | 47.6 | 35.6 | 53.3 | 61.6 | 58.9 | 52.1 | | 45.4 | 50.3 | |
| Asn GTT | | | | | 68.5 | 53.6 | 51.2 | 47.6 | | 51.8 | | | | |
| Lys TTT | | | | | | 57.5 | | 47.8 | | 62.1 | | 52 | | |
| Asp GTC | | | | | | | 61.6 | | | 61.8 | | | | |
| Glu TTC | | | | | | | | | | | | | | |
| Cys GCA | | | | | | | | | 54.8 | 31.5 | 24.5 | 45.4 | | |
| Trp CCA | | | | | | | | | | 16.4 | | | | |
| Arg ACG | | | | | | | | | | | | | | |
| Ser2 GCT | | | | | | | | | | | | 54.7 | | |
| Gly GCC | | | | | | | | | | | | | 72.6 | 69.2 |
| Gly TCC | | | | | | | | | | | | | | |

## Figure 3

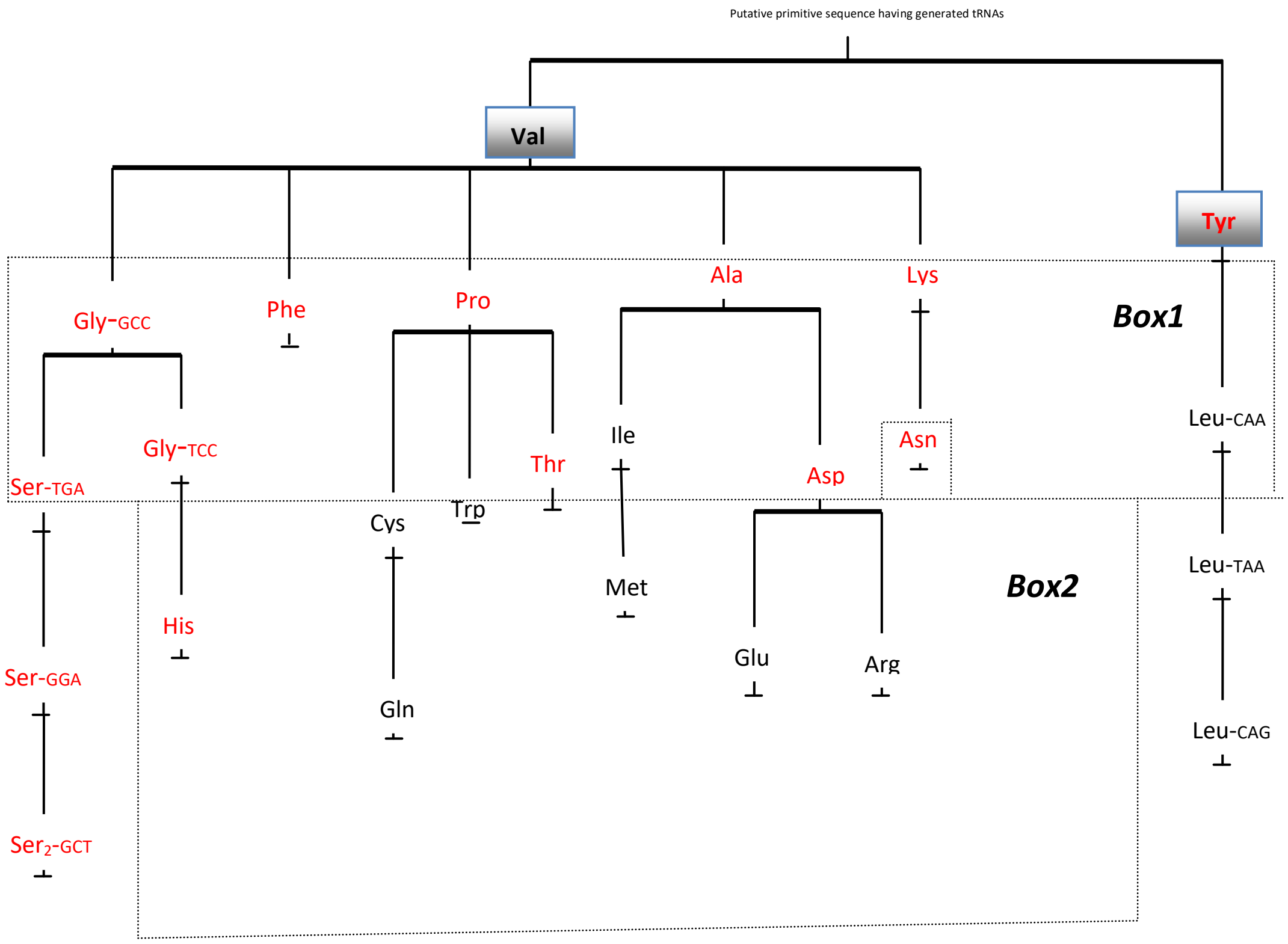

**Figure 4**

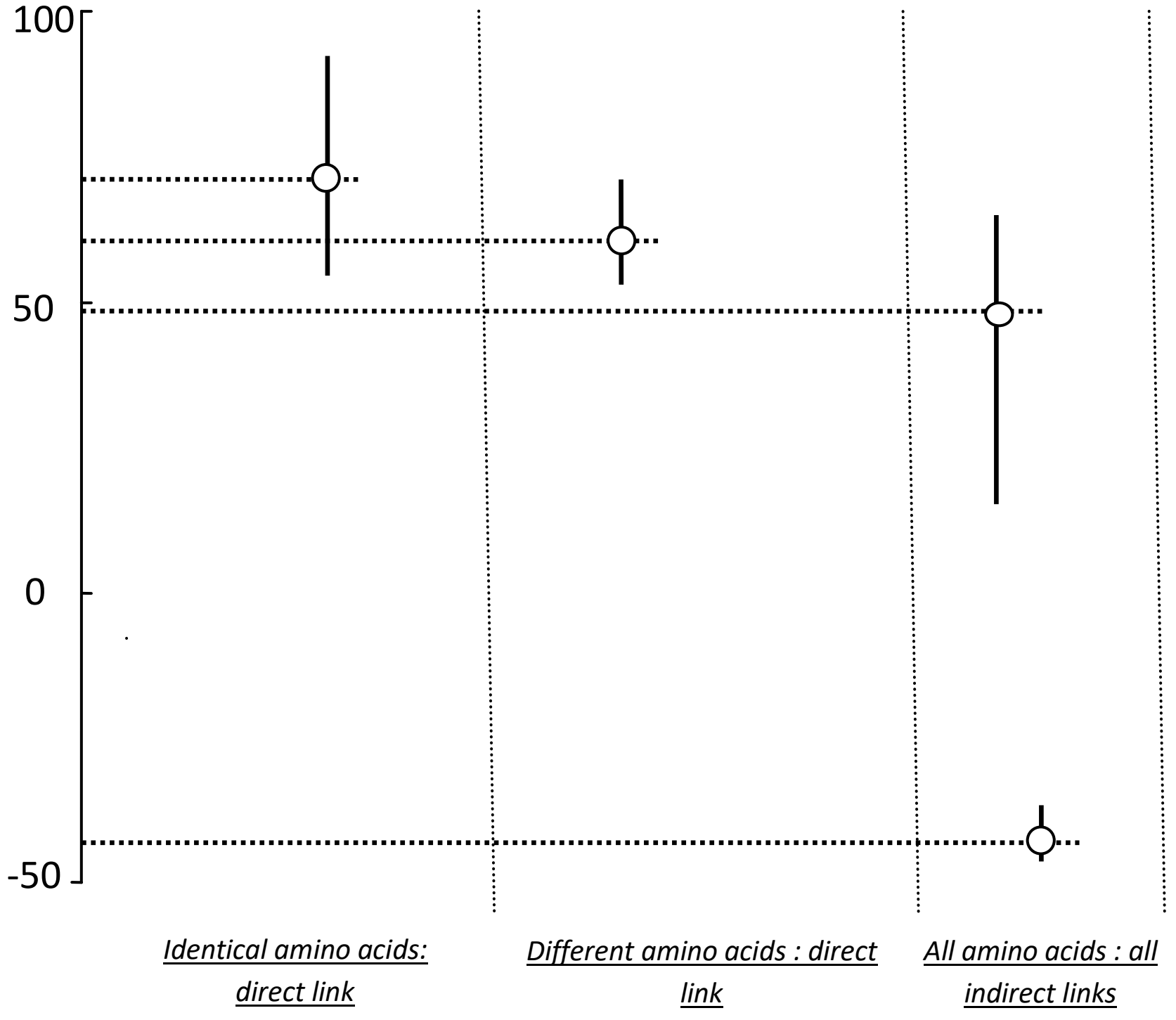
100
50
0
-50
Identical amino acids: direct link
Different amino acids : direct link
All amino acids : all indirect links

**Figure 5**

*( A A)* **G G G G G A G U A G C U C A G U U G G U A G A G C A C C U G C U U**

[**X X X ] A A G C A G G G G G U C G G C G G U U C G A A C C C G U C C C C C**

**U** *( U )*

**Figure 6** (a) (b)

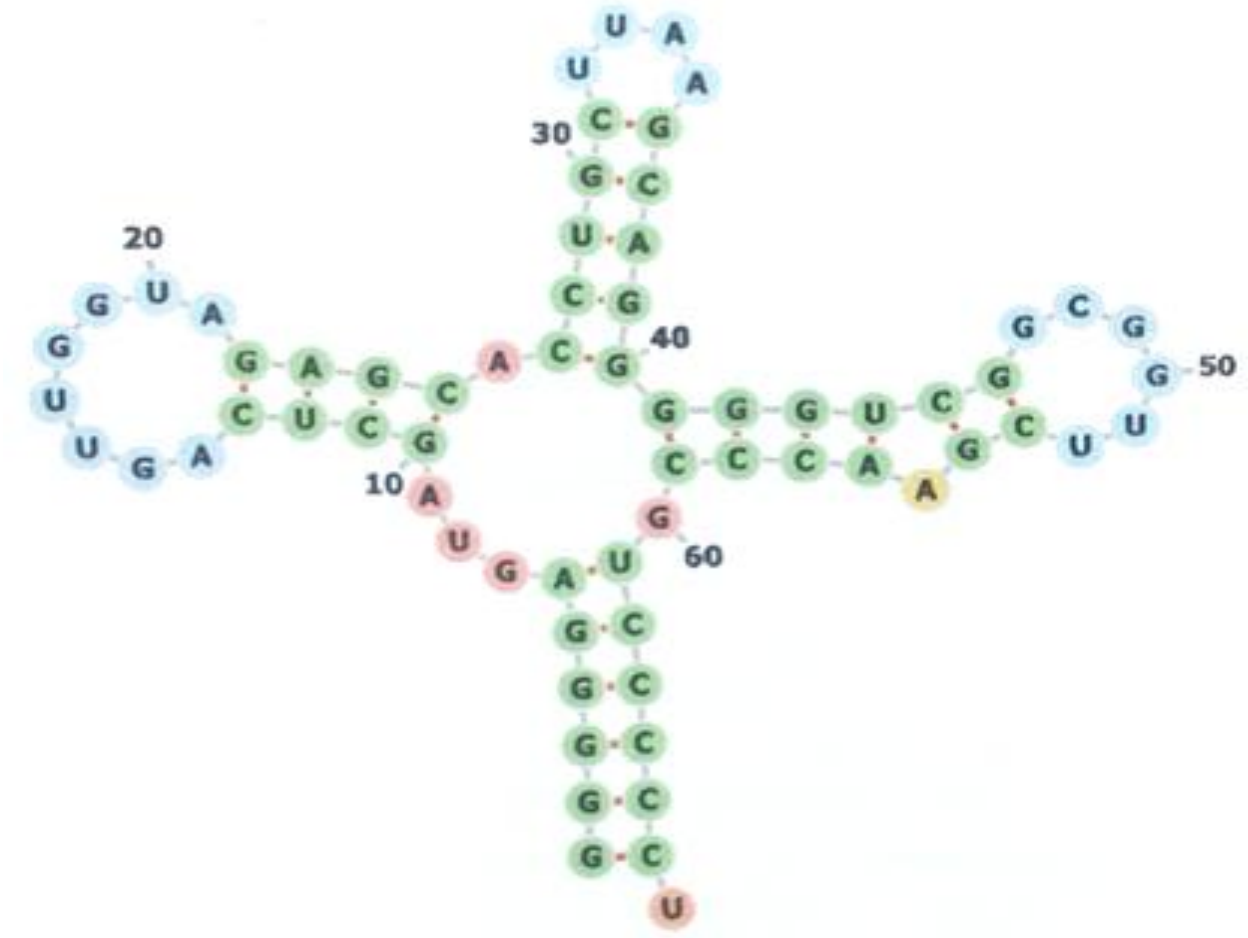

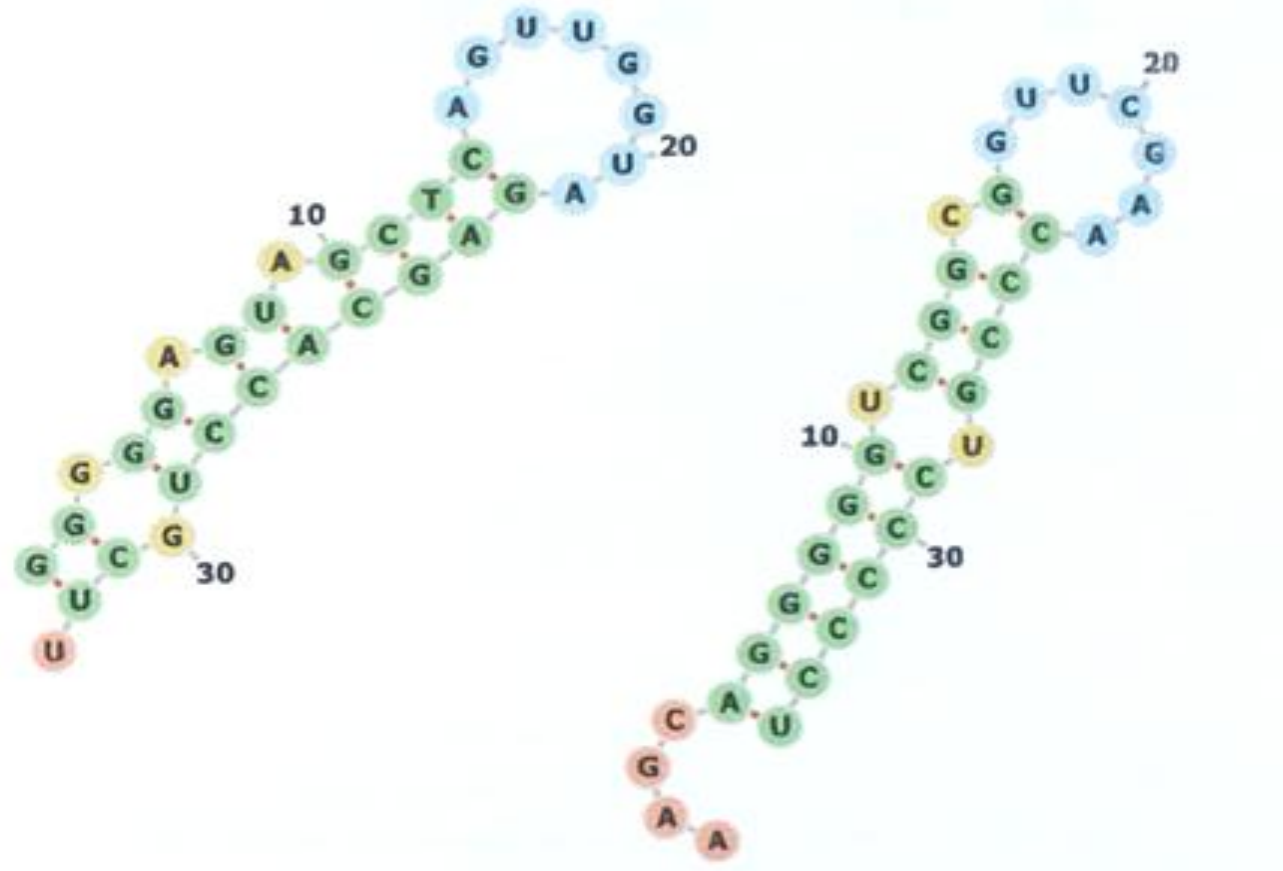

**Figure 7**

```
>--
 Waterman-Eggert score: 23;  8.3 bits; E(1) <  0.97
53.6% identity (53.6% similar) in 28 nt overlap (4-3

             10         20         30
unknow GGAGTAGCTCAGTTGGTAGAGCACCTGC
       : ::  : :: :  : : :: : : : :
unknow GCAGGGGGTCGGCGGTTCGAACCCGTCC
              10         20         30

>--
```

## Figure 8

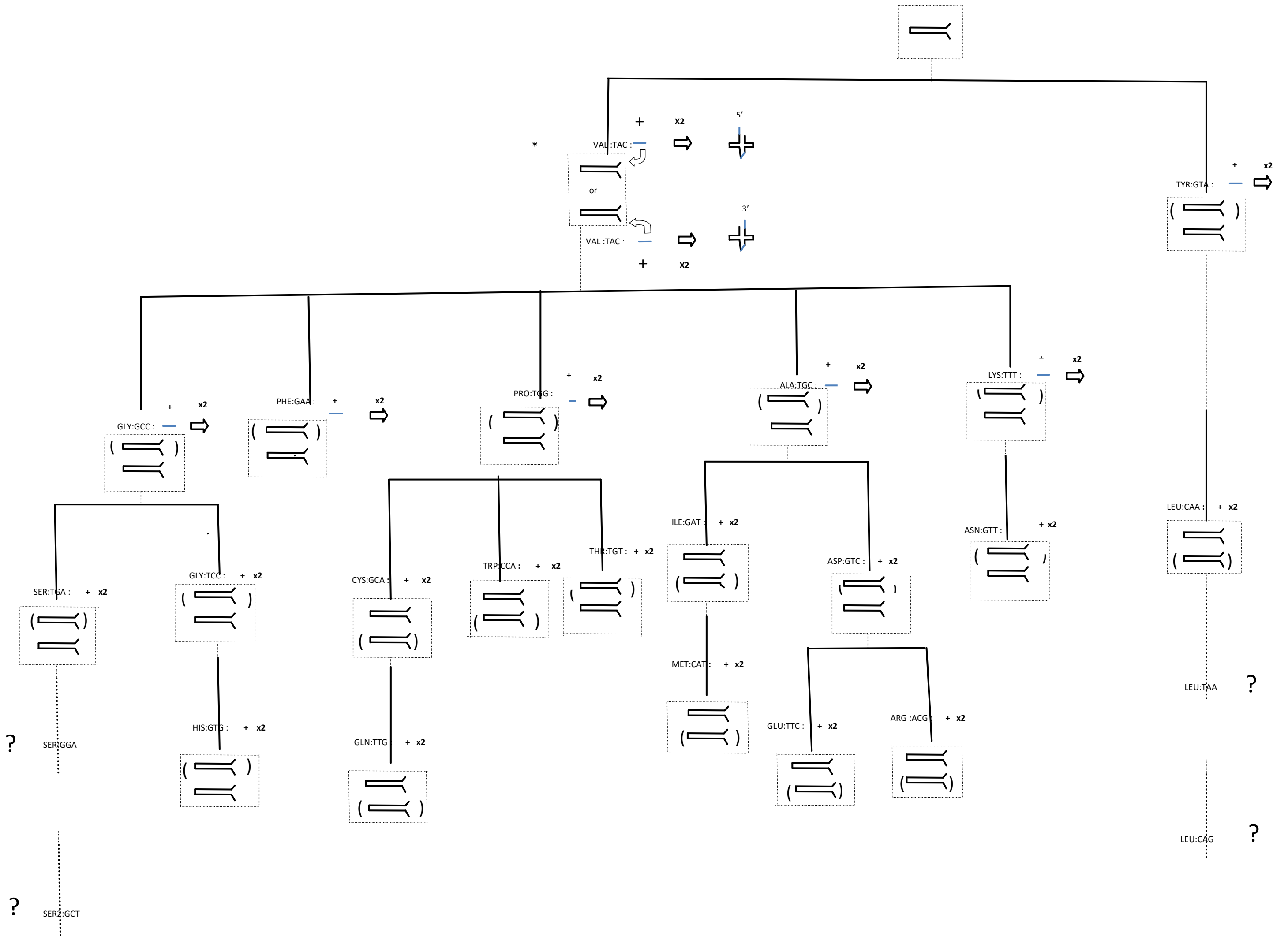

## Figure 9

| | Val GAC | LeuTAG | LeuGAG | Ala GGC | ThrGGT | Arg CCG | Arg TCT | Arg CCT |
|---|---|---|---|---|---|---|---|---|
| **Val TAC** | **68.5** | | | | | | | |
| **LeuCAA** | | **70.7** | **28-42.5** | | | | | |
| **LeuTAA** | | **67.4** | **69.3** | | | | | |
| **LeuCAG** | | **90.5** | **75.3** | | | | | |
| **LeuTAG** | | | **74.1** | | | | | |
| **Ala TGC** | | | | **81.1** | | | | |
| **Thr TGT** | | | | | **56.1** | | | |
| **ArgACG** | | | | | | **71** | **61.7** | **61** |
| **Arg CCG** | | | | | | | **65.8** | **71.1** |
| **Arg CCT** | | | | | | | **75.4** | |

**Figure 10**

## (a) : *E. coli/B. subtilis* tRNA comparison:

| | Val GAC *B. subtilis* | Ser GGA *B. subtilis* |
|---|---|---|
| **ValGAC1 *E. coli*** | 79.5 | |
| **ValGAC2 *E. coli*** | 74 | |
| **Val TAC *E. coli*** | 87.7 | |
| **Ser GGA *E. coli*** | | 76.4 |

## (b) : *E. coli/E. coli* tRNA comparison:

| | Gly CCC | Ser TGA | Ser CGA | Ser GCT | ProGGG | ProTGG | ThrGGT | Thr TGT | Gln CTG |
|---|---|---|---|---|---|---|---|---|---|
| **Gly GCC** | 68.5 | | | | | | | | |
| **Ser GGA** | | 50.6 | 70.1 | 78.9 | | | | | |
| **Pro GGG** | | | | | | 77 | | | |
| **Pro CGG** | | | | | 71.6 | 82.4 | | | |
| **Thr TGT** | | | | | | | 68.9 | | |
| **Thr CGT** | | | | | | | 71.2 | 72.6 | |
| **Gln TTG** | | | | | | | | | 90.3 |

**Figure 11**

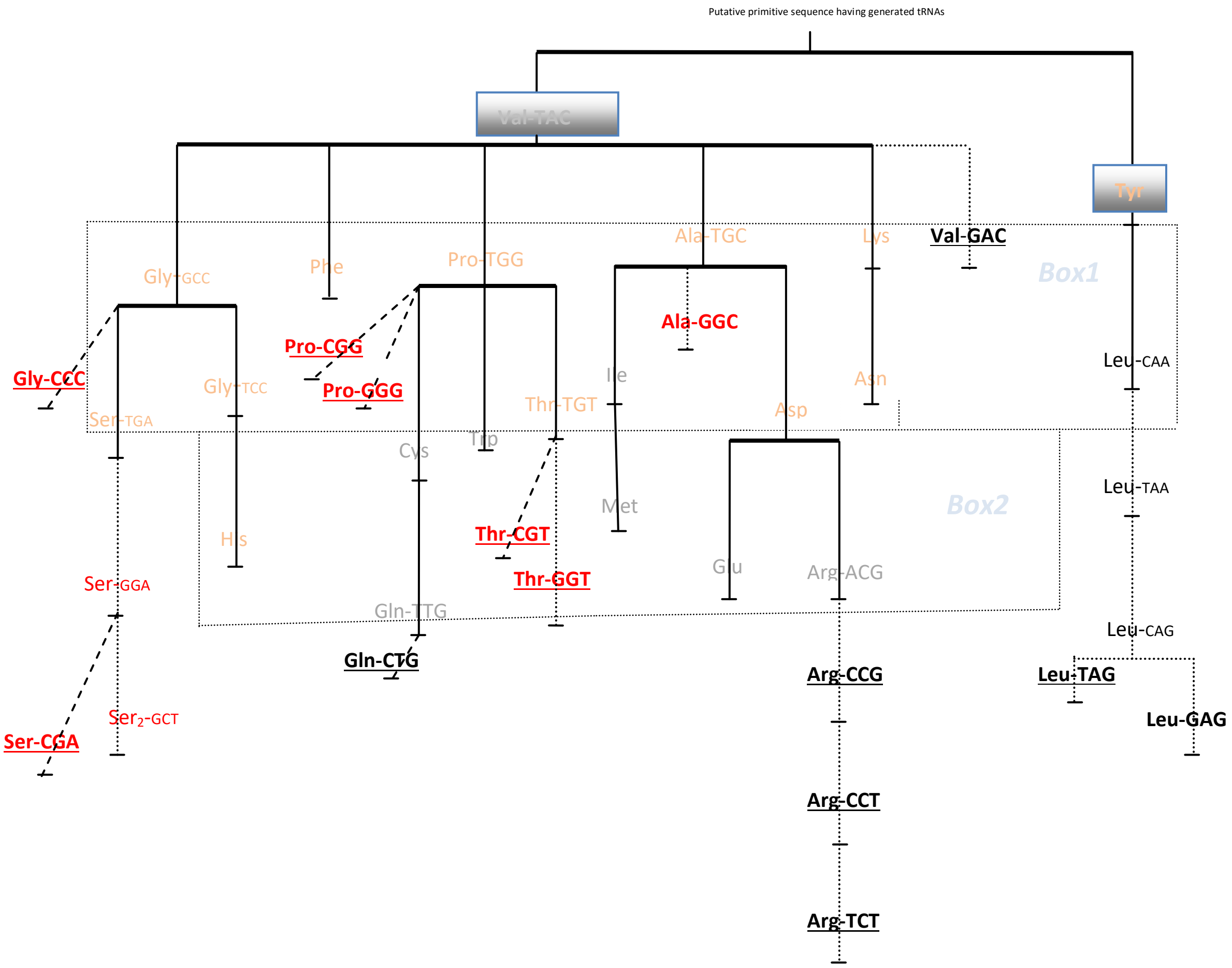

**Figure 12**

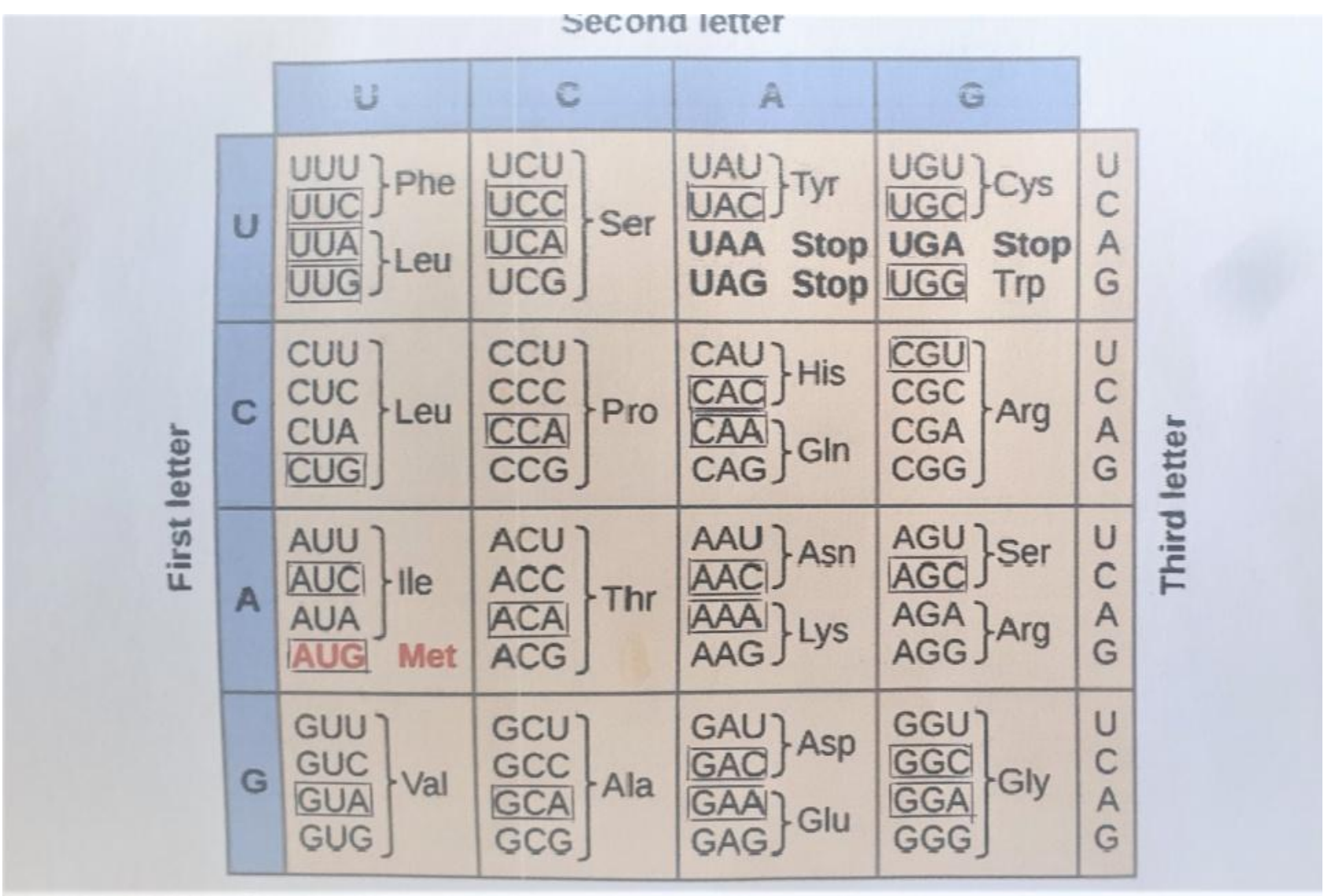

| First letter | Second letter: U | Second letter: C | Second letter: A | Second letter: G | Third letter |
|---|---|---|---|---|---|
| U | UUU Phe | UCU Ser | UAU Tyr | UGU Cys | U |
| U | UUC Phe | UCC Ser | UAC Tyr | UGC Cys | C |
| U | UUA Leu | UCA Ser | **UAA Stop** | **UGA Stop** | A |
| U | UUG Leu | UCG Ser | **UAG Stop** | UGG Trp | G |
| C | CUU Leu | CCU Pro | CAU His | CGU Arg | U |
| C | CUC Leu | CCC Pro | CAC His | CGC Arg | C |
| C | CUA Leu | CCA Pro | CAA Gln | CGA Arg | A |
| C | CUG Leu | CCG Pro | CAG Gln | CGG Arg | G |
| A | AUU Ile | ACU Thr | AAU Asn | AGU Ser | U |
| A | AUC Ile | ACC Thr | AAC Asn | AGC Ser | C |
| A | AUA Ile | ACA Thr | AAA Lys | AGA Arg | A |
| A | AUG Met | ACG Thr | AAG Lys | AGG Arg | G |
| G | GUU Val | GCU Ala | GAU Asp | GGU Gly | U |
| G | GUC Val | GCC Ala | GAC Asp | GGC Gly | C |
| G | GUA Val | GCA Ala | GAA Glu | GGA Gly | A |
| G | GUG Val | GCG Ala | GAG Glu | GGG Gly | G |